\documentclass{elsart}     

\input epsf

\begin{document}
\begin{frontmatter}
\title{A case for biotic morphogenesis of coniform stromatolites}
 
\author[ANUa]{M.T. Batchelor} 
\author[ANUb]{R.V. Burne} 
\author[UNSW]{B.I. Henry}
 and
\author[PO]{M.J. Jackson}
\address[ANUa]{Department of Theoretical Physics,\\ 
Research School of Physical Sciences and Engineering,\\
and Centre for Mathematics and its Applications, Mathematical Sciences Institute,\\
The Australian National University, Canberra ACT 0200, Australia}
\address[ANUb]{Department of Geology, The Australian National University,\\
Canberra ACT 0200, Australia}
\address[UNSW]{Department of Applied Mathematics, School of Mathematics,\\
The University of New South Wales, Sydney NSW 2052, Australia}
\address[PO]{PO Box 636, Mitchell ACT 2911, Australia}
 
\begin{abstract}
Mathematical models have recently been used to cast doubt on the biotic origin of stromatolites.  
Here by contrast we propose a biotic model for stromatolite morphogenesis which considers the 
relationship between upward growth of a phototropic or phototactic biofilm ($v$) and mineral 
accretion normal to the surface ($\lambda$). 
These processes are sufficient to account for the growth and form of many ancient stromatolities.  
Domical stromatolites form when $v$ is less than or comparable to $\lambda$.
Coniform structures with thickened apical zones, typical of Conophyton, form when $v >> \lambda$. 
More angular coniform structures, similar to the stromatolites claimed as the oldest macroscopic 
evidence of life, form when $v >>> \lambda$.
\end{abstract}


\end{frontmatter}

\section{Introduction}

Stromatolites preserve the only macroscopic evidence of life prior to the appearance of 
macro-algae \cite{K}.  
The biogenicity of stromatolites older than 3.2 Ga\footnote{1 Ga = $10^9$ years.} is 
unclear \cite{WBD,ARR,L,HGHT}.  
If they are indeed biotic, they are the oldest morphological evidence for life, 
now that the identification of 3.3 to 3.5 Ga microfossils \cite{SP} has been 
challenged \cite{B,G}.  
Here we propose a mathematical model for stromatolite morphogenesis that endorses a biotic origin 
for coniform stromatolites.  
It analyses interaction between upward growth of a phototropic or phototactic microbial mat and 
mineral accretion normal to the surface of the mat.  
Domical structures are formed when mineral accretion dominates.  
When vertical growth dominates, coniform structures evolve that reproduce the 
features of Conophyton, a stromatolite that flourished in certain low-sedimentation 
environments for much of the Proterozoic \cite{BSMP}.  
Increasing the dominance of vertical growth produces sharply-peaked conical forms, 
comparable to coniform stromatolites described from the 3.45 Ga Warrawoona Group, 
Western Australia \cite{HGHT}.

Some authors prefer to avoid a genetic definition for stromatolites \cite{GK}, 
but we regard them as laminated microbialites \cite{BM}, biomechanically \cite{W} 
and functionally \cite{KK} analogous to lithified sessile organisms, 
such as colonial corals, in which living tissue is restricted to the surface1 \cite{JLR}. 
In stromatolites the living tissue is a benthic microbial community (BMC) \cite{CR}. 
BMCs range from communities composed of a single species to complex trophic networks of 
photoautotrophs, chemoatotrophs and heterotrophs \cite{CR} in which species composition and 
diversity may change in response to environmental conditions \cite{BDFS}  
In cases where the BMC includes photoautotrophs, such as cyanobacteria, the stromatolite 
form represents a record of that community's response to light.
There have been very few previous attempts to model stromatolite morphogenesis mathematically. 
Verrecchia \cite{V} proposed a simulation model for microstromatolites in calcrete crusts using a 
diffusion-limited aggregation model \cite{WS}, but this is only relevant to modelling complexly 
branching stromatolites.  
Insights into morphogenesis of simpler forms may be gained using the interface evolution 
equation of Kardar, Parisi and Zhang (KPZ) \cite{KPZ} 
which contains parameters for surface-normal accretion, surface tension, and noise. 
Grotzinger and Rothman \cite{GR} attempted to simulate stromatolite form using 
a modified KPZ equation with explicit vertical growth.  
In their model vertical growth was considered to be due to the deposition of suspended sediment, 
surface-normal accretion was due to chemical precipitation, surface tension effects were related 
to both diffusive smoothing of the settled sediment and  chemical precipitation, and uncorrelated 
random noise represented surface heterogeneity and environmental fluctuations. 
Their model simulated the structure of a supposed stromatolite from the Cowles Lake Formation, 
Canada, and this led them to conclude that some and perhaps many stromatolites may be accounted 
for exclusively by abiotic processes. 
However, that model was subsequently modified to include a biotic process, mat growth, 
along with mineral precipitation in the surface-normal growth parameter \cite{GK}.  
As both mineral accretion and biological growth were linked in their surface-normal growth parameter, 
this model was unable to discriminate biotic effects. 
A rather different application of the KPZ equation has been proposed \cite{BBHW} 
in which the effects of 
microbial growth are included in the vertical growth parameter of the equation. 
This has been used to simulate the morphogenesis of stromatolites from Marion Lake, 
South Australia. 

\section{The model}

The biotic interpretation of fossil stromatolites is widely accepted, despite the fact that 
they rarely preserve any remains of the BMC which formed them \cite{GK}. 
As a result, attention has been focussed on how biotic stromatolites might be distinguished 
from abiotic accretions such as tufa, speleothems, and calcrete \cite{GK}. 
Several Proterozoic stromatolite forms grew in environments of low-sedimentation and their 
formation seems to have been due to the growth of a BMC, containing photosynthetic bacteria, 
and accretion of calcium carbonate in the resulting biofilm \cite{K,ARR,BSMP}. 
For forms which lack evidence of detrital material being trapped or bound by the BMC we 
propose a model for stromatolite morphogenesis which involves two processes only: 
\begin{enumerate}
\item upward growth of a phototropic or phototactic BMC, 
\item mineral accretion normal to the surface. 
\end{enumerate}
The function $h(x,t)$ represents the height of the profile above a horizontal baseline which 
evolves in time $t$ according to the equation
\begin{equation}
\frac{\partial h}{\partial t}= v 
+ \lambda \left(1+\frac{1}{2}\left(\frac{\partial h}
{\partial x}\right)^2\right).\label{eqn}
\end{equation}
The co-ordinate $x$ measures the distance along the baseline.
It is also equivalently a radial co-ordinate in the baseplane for circularly symmetric profiles. 
We interpret $v$ as the average rate of vertical growth due to photic response of microbes and 
$\lambda$ as the average rate of surface-normal growth due to mineral accretion.

\section{Results and discussion}

Although non-linear, equation (\ref{eqn}) can be solved with a change of variables using the 
method of characteristics \cite{BBHW} and prescribed initial profiles. 
The choice of initial profile is important. 
Cone-like initial profiles arise naturally in deformations of thin flat sheets \cite{LGLMW}. 
Fig. 1 shows examples of forms obtained from our model using initial profiles similar to those 
thought in field and laboratory studies to initiate coniform stromatolites \cite{WBB,H,BSMP,JMP}. 
The functional form of these solutions to equation (\ref{eqn}) is given by 
\begin{equation}
h(x,t)=\left\{\begin{array}{lr}
h_I(x,t), &
\qquad -\lambda t \, | \beta^-| - \gamma \le x \le -\lambda t \, |\beta^+|,\\
1 + (v + \lambda t) -\frac{x^2}{2\lambda t}, &\qquad 
-\lambda t \, |\beta^+| \le x \le \lambda t \, |\beta^+|,\\
h_I(-x,t),&
\qquad \lambda t \, |\beta^+| \le x \le \lambda t \, | \beta^-| + \gamma, 
\end{array}\right.
\label{sol}
\end{equation}
where 
\begin{equation}
h_I(x,t) = 1 + (v + \lambda t) + \frac{ a x^2 + \frac12 \lambda t |\beta^+|^2 + |\beta^+| x}
{1 - 2 a \lambda t},
\end{equation}
with $\beta^\pm = (a \pm \alpha^2)/\alpha$, $\gamma = \min(\alpha/|a|, 1/\alpha)$ and $\alpha^2 > |a|$.
The additional parameters, $\alpha$ and $a$, are used to tune the concavity or convexity of the flanks 
and the sharpness of the peak of the initial shape.

The results provide possible explanations for variations in coniform stromatolite morphogenesis. 
When $v$ is smaller than or comparable to $\lambda$ the result is a domical form (Figs. 1a, 1b and 1c). 
Coniform structures with thickened apical zones form when $v >>\lambda$ (Figs. 1d, 1e and 1f). 
More angular coniform structures form when $v >>> \lambda$ (Figs. 1g, 1h and 1i). 
The essential characterisitics of Conophyton \cite{M}, 
a columnar stromatolite composed of conical 
laminae with thickened crests (Figs. 2 and 3) are apparent in Figs. 1d, 1e and 1f. 
The laminae of Conophyton generally lack any evidence of trapping or binding of detrital sediment 
particles and the various microstructures that have been described all appear to result from a 
combination of BMC growth and carbonate precipitation \cite{KRS}. 
The form is recorded from Paleoproterozoic to Mezoproterozoic rocks world-wide, but becomes rare 
in the Neoproterozoic \cite{GK}. 
It flourished in extensive fields of conical columns up to 10m high in environments characterised 
by very low sedimentation rates \cite{D,H,BSMP,JMP}. 
The lack of evidence for significant sedimentation and evidence for an almost complete covering 
by a BMC during growth suggests that the Conophyton form is determined by two factors, 
light and mineral accretion. 
The thickening of the crestal zones in Figs. 1d, 1e and 1f is evocative of the thickened laminae 
and fenestrae in the delicate crestal zone of Conophyton \cite{BSMP}. 
Stromatolite fenestrae are voids in the lithified structure thought to have been left after the 
decay of the original BMC \cite{Monty}. 
A modern analogue for Conophyton has been recognised in hot-springs in 
Yellowstone National Park, USA \cite{WBB}, where it has been established 
that they form as a result of upward growth and motility of 
phototactic filamentous cyanobacteria combined with precipitation of silica, and that crestal 
fenestrae and thickening of laminae have been related to the preferential upward growth of the 
constructing microbes.

It has been concluded \cite{BDG} that crestal thickening of laminae and amplification of bedding 
irregularities are evidence for phototropic growth in stromatolites. 
The results of our model, equation (\ref{eqn}),  shown in Fig. 1,  together with field evidence, 
support the interpretation that the vertical growth parameter $v$ represents photic response of 
the BMC rather than sediment deposition. 
If the converse were true, coniform stromatolites would only form under conditions of high 
sedimentation which is precisely contrary to field evidence. 
Indeed, while sediment deposition would tend towards the smoothing of surface irregularities, 
growth due to photic response would tend to accentuate them. 
Our model shows that a combination of vertical phototropic or phototactic microbial growth and 
surface-normal mineral accretion can produce coniform forms and structures analogous to those 
found in both Archaean and Proterozoic coniform stromatolites. 
For example, there is a striking similarity between the model forms shown in Figs. 1g, 1h and 1i 
and the sharply-peaked coniform stromatolites in the Warrawoona Group \cite{HGHT}, 
thus supporting their 
biogenic origin and reinforcing the probability that photosynthetic microbes were components of 
Archaean BMCs.  
The various cases modelled in Figs. 1d, 1e and 1f can all be matched in Proterozoic Conophytons 
(Figs. 2 and 3). 
This sheds some light on why, after flourishing for much of the Proterozoic, Conophytons virtually 
disappeared in the Neoproterozoic \cite{KRS}. 
This demise has been linked to evolutionary changes in BMCs \cite{GK}, 
but since these would not have 
limited photic response, this seems untenable. 
Conophytons represent an effective growth strategy that is especially vulnerable to predation 
and competition \cite{JLR} and their demise is best explained as an indication of the evolution of 
greater biological diversity in the quiet marine environments that they had dominated for so long.

\ack
We thank Ian Hodgson and Patrick DeDeckker for discussions. 
This work has been supported by The Australian National University and the Australian Research Council.

\clearpage

\begin{figure}[ht!]
\vspace{120mm}
\includegraphics{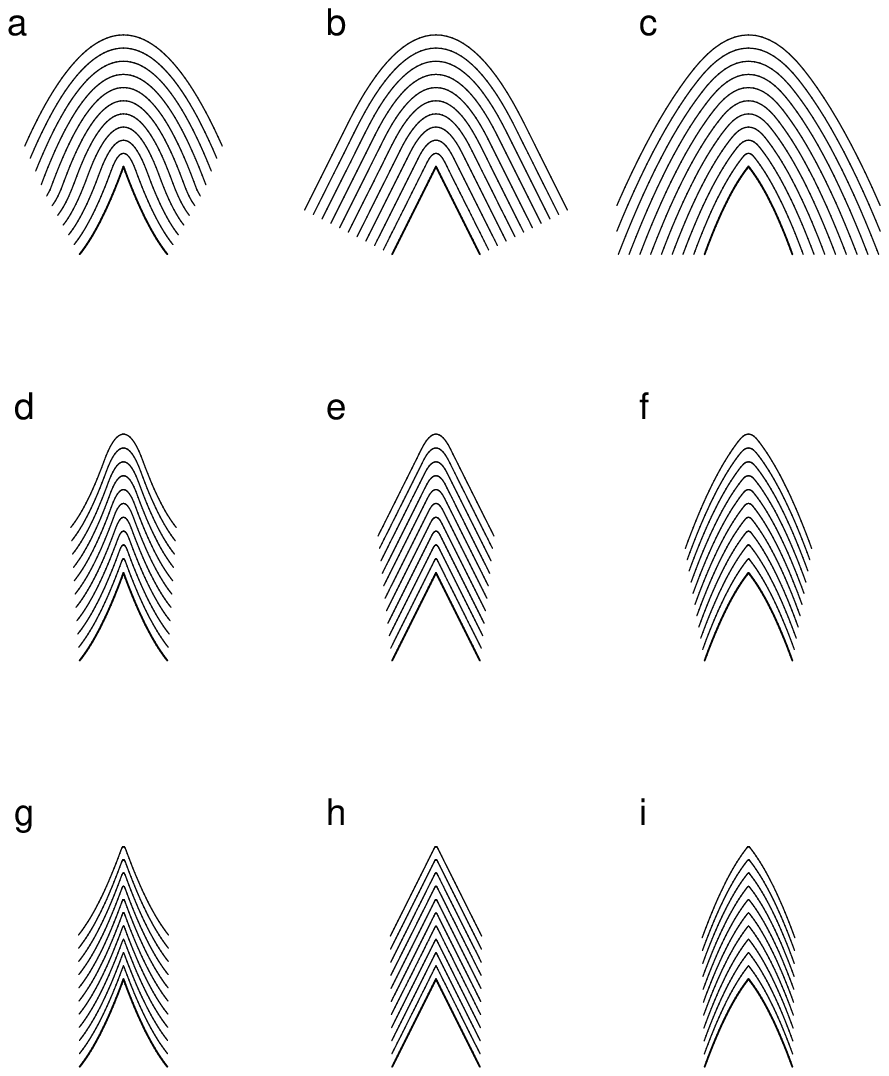}
\caption{Equal time snapshots of the surface height profile from the solution of the model, 
equations (1) and (2), for different values of the initial shape parameters $a$ and $\alpha$ 
and the growth parameters $\lambda$ and $v$. First column: $a = 1.5, \alpha = 2$, 
second column: $a = 0.01, \alpha = 2$, third column: $a = -1.5, \alpha = 2$. 
First row: $\lambda = 0.5, v = 1$, second row: $\lambda = 0.08, v = 1.5$, third row: 
$\lambda = 0.01, v = 1.5$.}
\end{figure}

\clearpage

\begin{figure}[ht!]
\vspace{170mm}
\includegraphics{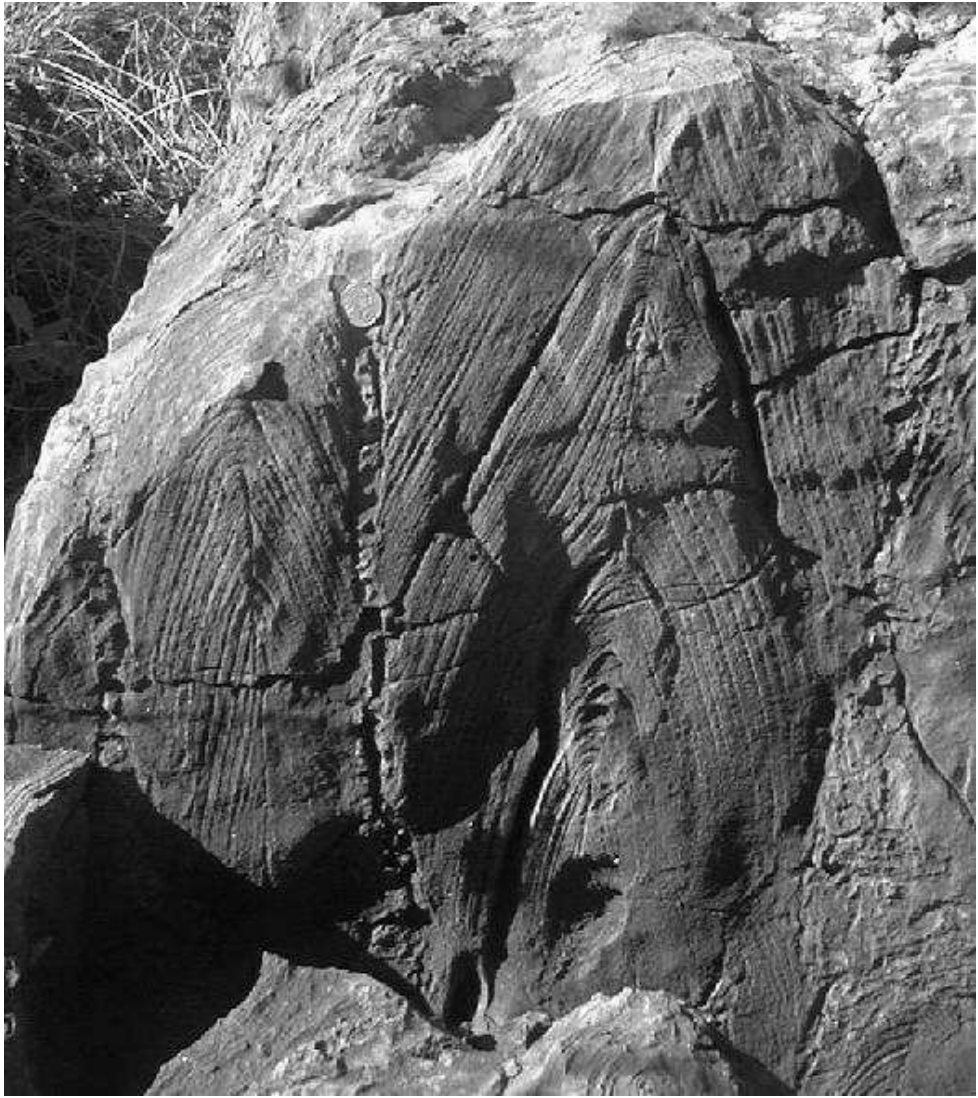}
\caption{Conophytons from the 1.7 Ga Dungaminnie Formation, NT, Australia 
(grid ref. 53K NB 0578973 8154545). Coin 28mm for scale. 
Compare the variation in conical laminations with Fig.1d, 1e and 1f.}
\end{figure}

\clearpage

\begin{figure}[ht!]
\vspace{180mm}
\includegraphics{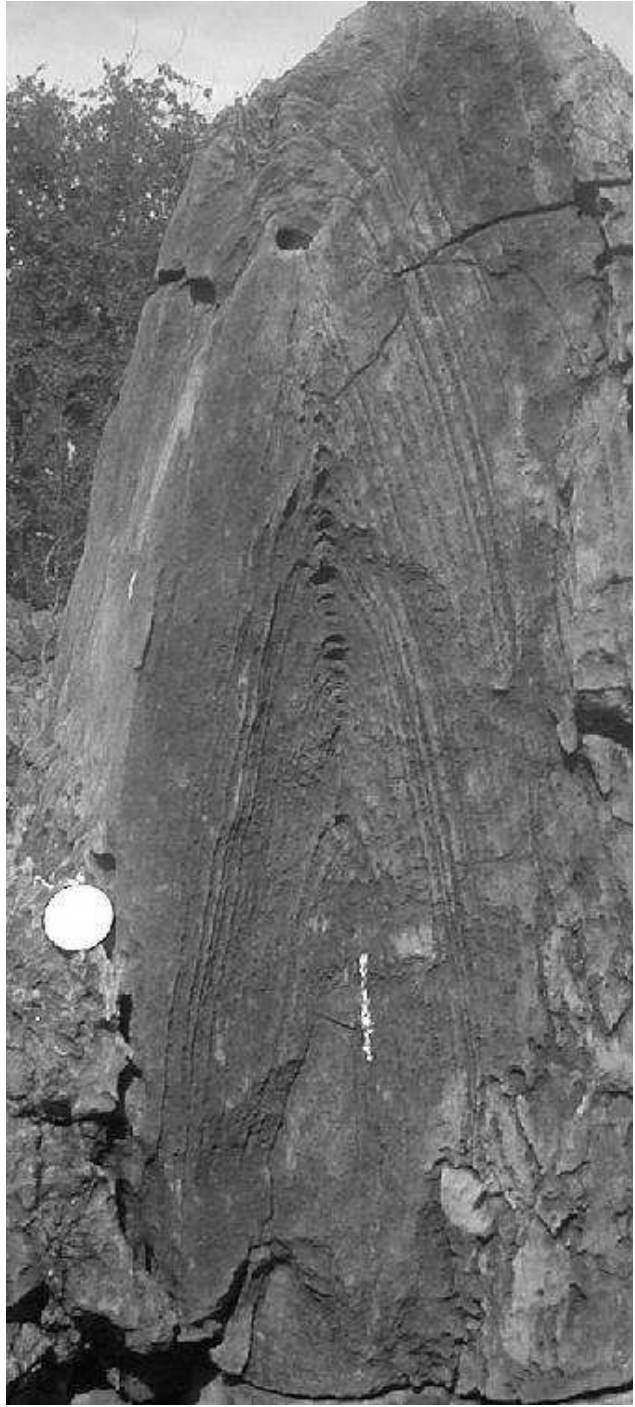}
\caption{Conophyton from the same locality as Fig. 2. Coin 28mm for scale. Note fenestrae, thought to have originally contained unlithified BMCs, and thickening of laminations in axial zone.}
\end{figure}

\end{document}